\documentclass[prb,twocolumn]{revtex4}
\usepackage{graphicx}

\newcommand{\sA}{{\scriptscriptstyle A}}
\newcommand{\sB}{{\scriptscriptstyle B}}
\newcommand{\sD}{{\scriptscriptstyle D}}

\newcommand{\sF}{{\scriptscriptstyle F}}
\newcommand{\sN}{{\scriptscriptstyle N}}
\newcommand{\sQ}{{\scriptscriptstyle Q}}
\newcommand{\sR}{{\scriptscriptstyle R}}
\newcommand{\sAL}{{\scriptscriptstyle \text{AL}}}

\newcommand{\sBCS}{{\scriptscriptstyle \text{BCS}}}

\begin{document}

\title{Superconducting fluctuations and the Nernst effect: A diagrammatic approach}

\author{Iddo Ussishkin}
\affiliation{Department of Physics, Princeton University, Princeton, NJ
08544}

\date{November 18, 2002}

\begin{abstract}
We calculate the contribution of superconducting fluctuations above the
critical temperature $T_c$ to the transverse thermoelectric response
$\alpha_{xy}$, the quantity central to the analysis of the Nernst
effect. The calculation is carried out within the microscopic picture
of BCS, and to linear order in magnetic field. We find that as $T
\rightarrow T_c$, the dominant contribution to $\alpha_{xy}$ arises
from the Aslamazov-Larkin diagrams, and is equal to the result
previously obtained from a stochastic time-dependent Ginzburg-Landau
equation [Ussishkin, Sondhi, and Huse, arXiv:cond-mat/0204484]. We
present an argument which establishes this correspondence for the heat
current. Other microscopic contributions, which generalize the
Maki-Thompson and density of states terms for the conductivity, are
less divergent as $T \rightarrow T_c$.
\end{abstract}

\maketitle

\section{Introduction and discussion of results}

In a superconductor, fluctuations of the superconducting order
parameter above the transition temperature $T_c$ affect various
properties such as the magnetic susceptibility and transport
coefficients. The study of superconducting fluctuations has a long
history (for reviews, see, e.g.,
Refs.~\onlinecite{Skocpol-Tinkham,Larkin-Varlamov}). More recently,
interest in fluctuation phenomena was renewed with the discovery of
high-temperature superconductors, where their short coherence lengths,
high critical temperatures, and layered structures imply a large regime
for the observation of fluctuations.~\cite{Fisher-Fisher-Huse}

One experiment that has aroused particular interest recently is that of
the Nernst effect: A temperature gradient $(- \nabla T) \parallel
\mathbf{\hat x}$ is applied in the presence of a magnetic field
$\mathbf{B} \parallel \mathbf{\hat z}$, and the electric field response
(in the absence of transport electric current) is measured in the
$\mathbf{\hat y}$ direction. Below $T_c$ in the vortex state the Nernst
effect is large due to vortex motion, while in the normal state it is
typically very small. In experiments in low temperature superconductors
no sign of superconducting fluctuations was reported as the temperature
was raised above $T_c$.~\cite{Huebener} In contrast, several different
experiments did observe the appearance of a fluctuation tail above the
critical temperature in the Nernst signal of the high-temperature
superconductors~\cite{Huebener-expt,Hagen-etal,Clayhold-etal,Hohn-etal}
(and also in the related Ettinghausen effect~\cite{Palstra-etal}). More
recently, the Nernst effect above $T_c$ has attracted considerable
attention with measurements showing a sizeable Nernst signal well above
$T_c$, in particular in the underdoped
regime.~\cite{Ong-expt,Capan-etal}

While a Nernst experiment is carried out under open circuit conditions,
the transport coefficients which arise naturally in a theoretical
description are those which relate the transport electric and heat
currents to the electric field and temperature gradient,
\begin{equation}\label{response-coefficients}
\left( \begin{array}{c} \mathbf{j}_{\text{tr}}^e \\ \\
\mathbf{j}_{\text{tr}}^\sQ \end{array} \right) = \left(
\begin{array}{ccc} \sigma & & \alpha  \\ \\
\tilde \alpha & & \kappa \end{array} \right) \left( \begin{array}{c}
\mathbf{E} \\ \\
- \nabla T \end{array} \right) \, .
\end{equation}
Here, $\sigma$ is the conductivity tensor, $\kappa$ a tensor of thermal
conductivity, and $\alpha$, $\tilde \alpha$ the thermoelectric tensors
(which obey the Onsager relations, $\tilde \alpha = T \alpha$).
Applying the open circuit condition to
Eq.~(\ref{response-coefficients}), the Nernst coefficient is expressed
in terms of the conductivity and thermoelectric tensors,
\begin{equation}\label{nernst}
\nu_\sN = \frac{E_y}{(-\nabla T) B} = \frac{1}{B} \, \frac{\alpha_{xy}
\sigma_{xx} - \alpha_{xx} \sigma_{xy}}{\sigma_{xx}^2 + \sigma_{xy}^2}
\, .
\end{equation}
The transverse thermoelectric response $\alpha_{xy}$, the quantity on
which this paper is focused, is of primary interest for understanding
the effect of superconducting fluctuations on the Nernst signal (as
discussed below).

In a recent paper, Ussishkin, Sondhi, and Huse discussed the
contribution of superconducting fluctuations to the thermoelectric and
thermal conductivity tensors using a stochastic time-dependent
Ginzburg-Landau equation (TDGL) in the limit of Gaussian
fluctuations.~\cite{Ussishkin-Sondhi-Huse} In this paper we revisit the
calculation of the transverse thermoelectric response $\alpha_{xy}$
using a diagrammatic calculation within BCS theory. The details of this
calculation are presented in subsequent sections. In the remainder of
this section we present and discuss the results of this paper.

We calculate $\alpha_{xy}$ above the critical temperature $T_c$, to
linear order in the magnetic field $\mathbf{B}
\parallel \mathbf{\hat z}$, and to leading order in $T - T_c$. We find
that, in two and three dimensions, the contribution of superconducting
fluctuations to the transverse thermoelectric response is
\begin{equation}\label{alpha-xy-result}
\alpha_{xy}^\sAL = \left\{
\begin{array}{lll}
\displaystyle \frac{1}{6 \pi} \, \frac{e}{\hbar} \,
\frac{\xi(T)^2}{\ell_\sB^2} \propto \frac{1}{T - T_c} & \qquad &
\text{for 2D,}
\\
\\
\displaystyle \frac{1}{12 \pi} \, \frac{e}{\hbar} \, \frac{\xi
(T)}{\ell_\sB^2} \propto \frac{1}{\sqrt{T - T_c}}& & \text{for 3D.}
\end{array}
\right. \end{equation} Here, $\ell_\sB = (\hbar c / e B)^{1/2}$ is the
magnetic length, and $\xi (T) \propto (T - T_c)^{-1/2}$ is the
coherence length of the superconducting order parameter.

It is well known (see, e.g.,
Refs.~\onlinecite{Skocpol-Tinkham,Larkin-Varlamov}) that
superconducting fluctuations enhance the conductivity above $T_c$ due
to both the Aslamazov-Larkin~\cite{Aslamazov-Larkin} and the
Maki-Thompson~\cite{Maki,Thompson} contributions (there are also
density of states terms, which are less important for the
conductivity). A similar identification of the microscopic
contributions applies to other transport coefficients. In the case of
the transverse thermoelectric response, as the superscript in
Eq.~(\ref{alpha-xy-result}) suggests, we find that the leading order
contribution to $\alpha_{xy}$ is due to the Aslamazov-Larkin diagrams
alone. The contribution of the Maki-Thompson and density of states
diagrams is less divergent as $T \rightarrow T_c$.

Physically, the Aslamazov-Larkin diagrams correspond to the
contribution of thermal fluctuations of the order parameter. Their
contribution to $\alpha_{xy}$ may be viewed either as the transport
heat current carried by such fluctuations when they respond to an
electric field, or as the transport electric current carried by the
fluctuations as they respond to a temperature gradient (all in the
presence of the magnetic field). The same physics is identically
described by the Gaussian approximation to a stochastic TDGL. (We will
have more to remark on the correspondence between the two approaches in
subsequent sections.) Indeed, the result obtained in this paper,
Eq.~(\ref{alpha-xy-result}), is identical to the result obtained in the
Gaussian approximation to the stochastic TDGL in
Ref.~\onlinecite{Ussishkin-Sondhi-Huse}.

Before discussing the experimental situation, we comment on two
assumptions made in this paper. First, we assume that the order
parameter has $s$-wave symmetry. In the context of the high-temperature
superconductors it is of interest to consider also the case of $d$-wave
symmetry in this approach. We note here that this will not affect the
conclusions of this paper: the results of the stochastic TDGL would
still correspond to the Aslamazov-Larkin contribution; and the
arguments showing that Maki-Thompson and density of states terms are
less divergent remain valid in this case as well.~\cite{Yip} (We do not
consider here the related issue, of whether nodal quasiparticles, which
appear when the temperature is lowered and the condensate is formed,
contribute to $\alpha_{xy}$.) Second, we assume particle-hole symmetry
(i.e., neglecting any contributions which arise due to asymmetry around
the Fermi surface in properties such as the density of states). For
quantities that do not vanish in this limit this is a very good
approximation for a BCS superconductor. (On the other hand,
particle-hole symmetry implies that $\sigma_{xy} = \alpha_{xx} =
\kappa_{xy} = 0$, and therefore in calculating these transport
coefficients it is necessary to break this symmetry.) We note that the
contribution of the normal metallic state to $\alpha_{xy}$ also
vanishes in this limit. However, this is not required by symmetry, and
indeed is no longer the case once superconductivity is taken into
account.

We now return to the discussion of the Nernst coefficient $\nu_\sN$. As
noted in Eq.~(\ref{nernst}), $\nu_\sN$ is related to both the
conductivity and thermoelectric tensors. However, the main effect of
superconducting fluctuations on the Nernst signal above $T_c$ is due to
$\alpha_{xy}$. Indeed, the contribution of fluctuations to the second
term in the numerator of Eq.~(\ref{nernst}) is small due to
considerations of particle-hole symmetry. Moreover, not too close to
$T_c$ the conductivity is dominated by the normal state contribution.
It follows that the main contribution of superconducting fluctuations
to the Nernst signal (to linear order in $B$) is $\alpha_{xy}^\sAL /
\sigma_{xx}$, with $\sigma_{xx}$ being the normal state contribution.

Since the result for $\alpha_{xy}^\sAL$, Eq.~(\ref{alpha-xy-result}),
depends only on the coherence length $\xi(T)$, and in a simple manner,
comparison with experiment should apparently be straightforward. In
Ref.~\onlinecite{Ussishkin-Sondhi-Huse}, such a comparison for a
high-temperature superconductor was presented. On the other hand, in
low-temperature superconductors, for which BCS theory is certainly
applicable, the appearance of the fluctuation tail in the Nernst signal
was not previously reported to the best of our
knowledge.~\cite{Huebener} The reason for this is that low-temperature
superconductors are typically also good conductors in the normal state.
Consequently $\alpha_{xy}^\sAL / \sigma_{xx}$, the contribution of
superconducting fluctuations to the Nernst signal, is strongly
suppressed in bulk low-temperature superconductors.

The situation can be improved considerably by looking at a thin film,
which is effectively a two dimensional superconductor if the coherence
length is larger than the film thickness. First, the fluctuation tail
of $\alpha_{xy}$ is enhanced by going to lower dimensionality, as is
evident in Eq.~(\ref{alpha-xy-result}). [The result for two dimensions
in Eq.~(\ref{alpha-xy-result}) is to be divided by the film thickness
to obtain the result for a thin film.] Second, such films may have a
significantly lower normal state conductivity. Taken together, these
effects may considerably enhance the contribution of fluctuations to
the Nernst signal. A similar situation may occur in a layered structure
with weak coupling between the layers and with a low normal-state
conductivity, as is the case for the high-temperature
superconductors.~\cite{Ussishkin-Sondhi-Huse}

In the remainder of the paper we present the details of our
calculation. In Sec.~\ref{sec-for} we discuss the Kubo formula for
$\alpha_{xy}$, and discuss an important aspect of the problem, namely
the role of \emph{bulk magnetization currents} and the proper
subtraction of their contribution.~\cite{Cooper-Halperin-Ruzin} We also
briefly present known results for the propagator of superconducting
fluctuations. The diagrams that appear in the calculation of
$\alpha_{xy}$, and the physics they describe, are discussed in
Sec.~\ref{sec-diagrams}. In Sec.~\ref{sec-j} we discuss the calculation
of the heat current vertex. A general argument regarding its
calculation is given, establishing the correspondence to the heat
current in the TDGL. In Sec.~\ref{sec-al} we calculate the
Aslamazov-Larkin diagrams, and obtain Eq.~(\ref{alpha-xy-result}). The
Maki-Thompson and density of states diagrams are considered in
Sec.~\ref{sec-mt}. Finally, we summarize our discussion in
Sec.~\ref{sec-summary}.

\section{Formalism}
\label{sec-for}

In this section we present a few results which will form the basis of
the calculations in the sections that follow. In Sec.~\ref{sec-Kubo} we
discuss linear response theory for the transverse thermoelectric
response. We present the Kubo formula for $\alpha_{xy}$, and discuss
the subtraction of \emph{bulk magnetization currents}. Presented here
only for the sake of completeness, Sec.~\ref{sec-fl} briefly discusses
known results for the propagator of superconducting fluctuations.

\subsection{Kubo formula for $\alpha_{xy}$}
\label{sec-Kubo}

In this paper the thermoelectric tensor is considered by calculating
the heat current response to an electric field. Alternatively, one
could consider within linear response theory the electric current
response to Luttinger's ``gravitational field''.\cite{Luttinger} The
result for $\alpha_{xy}$ is of course independent of which formulation
is used, and results are presented in terms of one of them only for
convenience.

The heat current response to an electric field is related to the heat
current-electric current correlator by using the standard Kubo formula.
Here, we are interested in the calculation of this response to linear
order in the magnetic field. This amounts to introducing an additional
current vertex coupled to the magnetic field.~\cite{Altshuler-Aronov}
Here we present this result as the Kubo formula for the linear response
to both an electric and magnetic field. The electric field $\mathbf{E}
\parallel \mathbf{\hat x}$ and the magnetic field $\mathbf{B} \parallel
\mathbf{\hat z}$ are introduced at finite frequency and wavevector,
respectively, using the vector potential
\begin{equation}
\mathbf{A} =  \frac{c \mathbf{E}}{i \Omega} e^{-i \Omega t} + \frac{B
\mathbf{\hat y}}{i Q} e^{i Q \mathbf{\hat x} \cdot \mathbf{r}} .
\end{equation}
The heat current in the $\mathbf{\hat y}$ direction, in response to the
electric and magnetic fields (in the d.c.\ limit), is given by
\begin{equation}\label{Kubo-jyQ-EB}
\frac{j^\sQ_y}{E B} = - \lim_{\Omega,Q \rightarrow 0} \frac{1}{\Omega Q
c} \text{Re} \left. \left[ \Lambda (Q, \Omega_m) \right] \right|_{i
\Omega_m \rightarrow \Omega + i 0}.
\end{equation}
Here, the three current correlator $\Lambda$ is defined by
\begin{eqnarray}\label{jjj-correlator}
\Lambda (Q, \Omega_m) & = & - \int_0^\beta d \tau \, d \tau' \, e^{i
\Omega_m \tau} \int d \mathbf{r} \, d \mathbf{r'} \, e^{i Q
\mathbf{\hat x} \cdot (\mathbf{r'} - \mathbf{r})} \nonumber
\\
& & \qquad \times \, \left\langle T_\tau j^\sQ_y (\mathbf{r}, \tau)
j^e_y (\mathbf{r'}, \tau') j^e_x (0) \right\rangle ,
\end{eqnarray}
where $\Omega_m = 2 \pi m T$ is a bosonic Matsubara frequency (with
units in which $\hbar = k_\sB = 1$), and the upper limit of integration
over imaginary times $\tau$ and $\tau'$ is the inverse temperature
$\beta = 1 / T$. In Eq.~(\ref{Kubo-jyQ-EB}), an analytic continuation
of $\Omega_m$ to real frequencies is performed before the zero
frequency limit is taken.

An important aspect of the calculation of the transverse thermoelectric
response, discussed in detail by Cooper, Halperin, and
Ruzin,~\cite{Cooper-Halperin-Ruzin} is the need to account for
\emph{bulk magnetization currents}.
This issue arises because the
microscopic electric and heat currents, as calculated by the Kubo
formula, are composed of transport and magnetization currents,
\begin{equation}
\mathbf{j}^e = \mathbf{j}_{\text{tr}}^e + \mathbf{j}_{\text{mag}}^e \,
, \qquad \mathbf{j}^\sQ = \mathbf{j}^\sQ_{\text{tr}} +
\mathbf{j}^\sQ_{\text{mag}} \, .
\end{equation}
The magnetization currents are currents that circulate in the sample
and do not contribute to the net currents which are measured in a
transport experiment. On the other hand, they do contribute to the
total microscopic currents, and it is thus necessary to subtract them
from the total currents to obtain the transport current response. In
the presence of an applied electric field, it was shown in
Ref.~\onlinecite{Cooper-Halperin-Ruzin} that the magnetization current
is given by
\begin{equation}
\mathbf{j}_{\text{mag}}^\sQ = c \mathbf{M} \times \mathbf{E} ,
\end{equation}
where $\mathbf{M}$ is the equilibrium magnetization (in the absence of
the electric field). It then follows that the transverse thermoelectric
response is given by
\begin{equation}
\alpha_{yx} = - \alpha_{xy} = \frac{j_y^\sQ}{E} - c M_z,
\end{equation}
where $j_y^\sQ / E$ is found using the Kubo formula,
Eqs.~(\ref{Kubo-jyQ-EB}) and~(\ref{jjj-correlator}).

In calculating the electric current response to a ``gravitational
field'' $\psi$, a similar situation arises, where in order to obtain
the transport current the electric magnetization current has to be
subtracted. The latter is given in this case
by~\cite{Cooper-Halperin-Ruzin}
\begin{equation}
\mathbf{j}^e_{\text{mag}} = - c \mathbf{M} \times \nabla \psi .
\end{equation}
In Ref.~\onlinecite{Ussishkin-Sondhi-Huse} the total heat current
response to an electric field and the total current response to a
temperature gradient were calculated using the stochastic TDGL. The
apparent discrepancy between the results for the total currents and the
Onsager relations was invoked to demonstrate the need of subtracting
out the magnetization currents. In contrast, in the linear response
formalism the calculation for the electric current yields its response
to a ``gravitational field'' gradient, and this apparent discrepancy
does not arise. The magnetization currents (and the total currents)
trivially obey the same Onsager relations obeyed by the transport
currents. Nevertheless, the magnetization currents must be subtracted
to obtain the correct result for the transport coefficients.

\subsection{Fluctuation propagator}
\label{sec-fl}

The contribution of superconducting fluctuations to the current
correlator in Eq.~(\ref{jjj-correlator}) is calculated in this paper
for a BCS superconductor (with s-wave symmetry), with Hamiltonian
\begin{eqnarray}\label{Hamiltonian}
\mathcal{H} & = & \sum_\mathbf{k, \sigma} \epsilon_\mathbf{k} \,
c^\dagger_\mathbf{k, \sigma} c_\mathbf{k, \sigma} + \sum_{\mathbf{k},
\mathbf{q}, \sigma} U_{\mathbf{q}} \, c^\dagger_{\mathbf{k} +
\mathbf{q}, \sigma} c_{\mathbf{k}, \sigma} \nonumber
\\
& & + \, \lambda \sum_{\mathbf{k}, \mathbf{k'}, \mathbf{q}}
c^\dagger_{\mathbf{k'}, \uparrow} c^\dagger_{-\mathbf{k'} + \mathbf{q},
\downarrow} c_{-\mathbf{k} + \mathbf{q}, \downarrow} c_{\mathbf{k},
\uparrow}.
\end{eqnarray}
Here, $\epsilon_\mathbf{k} = k^2 / 2 m$ is the kinetic energy of the
electrons, $\sigma = \, \uparrow, \downarrow$ is their spin,
$U_\mathbf{q}$ is the disorder potential (with the usual Gaussian
distribution), and $\lambda < 0$ is the attractive BCS interaction
(where only states with energy differing from the Fermi energy by at
most $\omega_\sD$ participate in the interaction term). The relevant
diagrams for superconducting fluctuations are calculated using the
finite-temperature diagrammatic technique. This approach is analogous
to the one used, e.g., in the case of the conductivity, leading to the
Aslamazov-Larkin~\cite{Aslamazov-Larkin} and
Maki-Thompson~\cite{Maki,Thompson} contributions. (For a detailed
account of the diagrammatic calculation of the conductivity, see, e.g.,
Ref.~\onlinecite{Larkin-Varlamov}.) A basic ingredient in this
calculation is $L$, the propagator of superconducting fluctuations.

\begin{figure}
\begin{center}
  \includegraphics[width=2in]{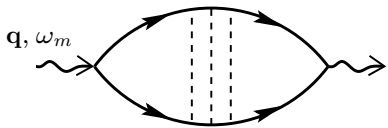}
\end{center}
\caption{\label{fig-pi} Diagram for the non-interacting two particle
propagator $\Pi (\mathbf{q}, \omega_m)$. Lines with arrows are
electronic Green functions, and disorder is shown in the ladder
approximation by dashed lines.}
\end{figure}

Accounting for the electron-electron interaction in the ladder
approximation, the fluctuation propagator $L$ is related to the
non-interacting two-particle propagator $\Pi$ through
\begin{equation}
L (\mathbf{q}, \omega_m) = \left[ \lambda^{-1} - \Pi (\mathbf{q},
\omega_m) \right]^{-1}
\end{equation}
(see Fig.~\ref{fig-pi} for the diagram of $\Pi$). To obtain the
retarded fluctuation propagator, the Matsubara frequency is
analytically continued to the real axis ($i \omega_m \rightarrow \omega
+ i 0$) and $\Pi$ is calculated to leading order in $\mathbf{q}$ and
$\omega$. Assuming particle-hole symmetry, the retarded fluctuation
propagator is then
\begin{equation}\label{L}
L^\sR (\mathbf{q}, \omega) = - \frac{1}{\nu} \, \frac{1}{\epsilon +
\eta q^2 - i \omega \tau_\sBCS}
\end{equation}
(and the advanced fluctuation propagator is $L^\sA (\mathbf{q}, \omega)
= [L^\sR (\mathbf{q}, \omega)]^*$). In Eq.~(\ref{L}), $\nu$ is the
density of states per spin, $\epsilon = \ln (T / T_c) \approx (T - T_c)
/ T_c$ is the reduced temperature, $\tau_\sBCS = \pi / 8 T_c$, and
\begin{eqnarray}
\lefteqn{\eta = - D \tau_{\text{el}}} \\ \nonumber & \times & \left[
\psi \left( \frac{1}{2} + \frac{1}{4 \pi T_c \tau_{\text{el}}} \right)
- \psi \left( \frac{1}{2} \right) - \frac{1}{4 \pi T_c
\tau_{\text{el}}} \psi' \left( \frac{1}{2} \right) \right] .
\end{eqnarray}
Here, $\tau_{\text{el}}$ is the elastic scattering time, $D = v_\sF^2
\tau_{\text{el}} / d$ is the diffusion constant (for $d$ dimensions),
and $\psi(x)$ is the digamma function. The parameters in Eq.~(\ref{L})
are directly related to the coefficients appearing in a TDGL for the
order parameter. In particular, $\xi (T) = \sqrt{\eta / \epsilon}$ is
the superconducting coherence length, and $\tau_\sBCS$ is the
relaxation time for the order parameter fluctuations.

\section{Diagrams and interpretation}
\label{sec-diagrams}

\begin{figure*}
\begin{center}
  \includegraphics[width=7in]{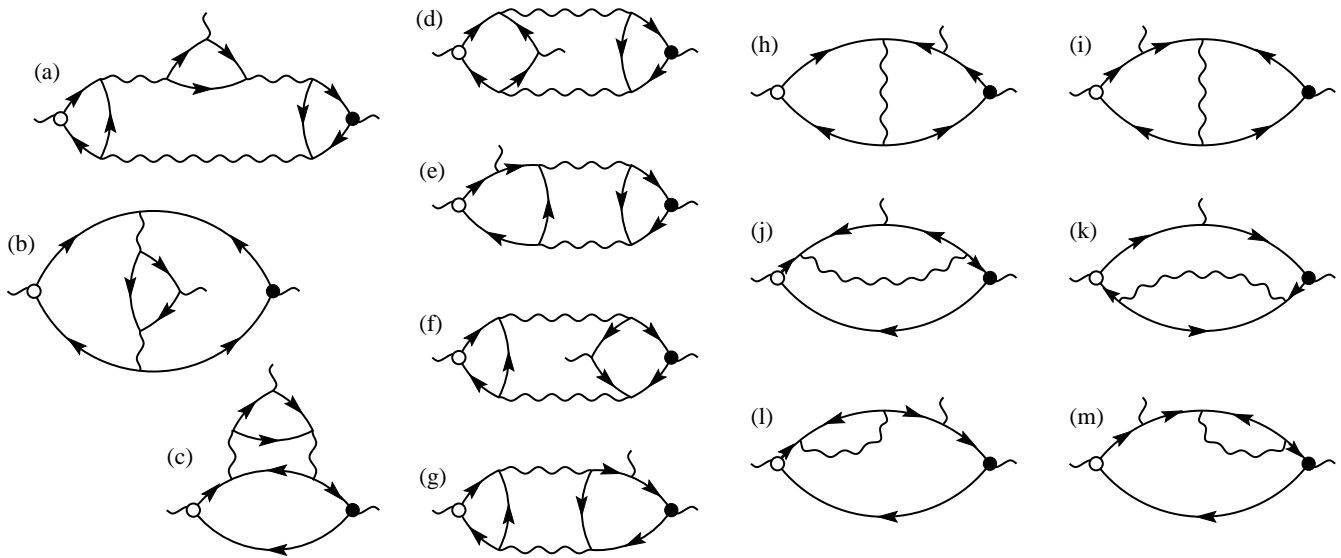}
\end{center}
\caption{\label{fig-diagrams} Diagrams arising in the calculation of
the transverse thermoelectric response [see
Eqs.~(\ref{Kubo-jyQ-EB})--(\ref{jjj-correlator})]. A wavy line
represents the fluctuation propagator, and the lines with arrows are
the electronic Green functions. The vertices are the heat current
vertex $j_y^\sQ $ (open circle), the electric current vertex coupled to
the electric field $j^e_x$ (full circle), and the electric current
vertex coupled to the magnetic field $j^e_y$ (no circle). ``Mirror
image'' diagrams, which may be obtained by reversing the arrow
direction on each of the electronic Green functions in the diagrams
above [except diagrams (b), (d), and (f), for which the ``mirror
image'' is not a new diagram], are not presented [See Fig.~\ref{fig-al}
below for the ``mirror image'' of diagram (a)]. Also not presented are
the different possibilities of adding disorder to each of these
diagrams.}
\end{figure*}

In this section we present the diagrams which appear in the calculation
of the correlator (\ref{jjj-correlator}), and discuss the physical
processes which they represent. For this purpose, the microscopic
picture is perhaps best recast in terms of a quantum functional
integral approach.~\cite{Svidzinskii,Popov} We begin this section by
briefly discussing this approach.

The expectation value of a current operator $\mathbf{j}$ (which can be
either the electric current or the heat current) in response to a
driving field $\phi$ may be expressed in terms of an imaginary time
functional integral
\begin{equation}
\langle \, \mathbf{j} \, \rangle = \frac{\int D \psi \, D \bar \psi \,
\mathbf{j} \, e^{-S [\psi, \bar \psi, \phi]}}{\int D \psi \, D \bar
\psi \, e^{-S [\psi, \bar \psi, \phi]}}.
\end{equation}
Here, $\psi$ and $\bar \psi$ are the fermion fields, and the action $S$
is given by
\begin{equation}
S = \int_0^\beta d \tau \int d \mathbf{x} \left[ \sum_\sigma \bar
\psi_\sigma (x) \partial_\tau \psi_\sigma (x) + \mathcal{H} (x)
\right],
\end{equation}
where $\beta$ is the inverse temperature, $x = (\mathbf{x}, \tau)$, and
$\mathcal{H}$ is the Hamiltonian density. Introducing a pairing field
$\Delta$ via the usual Hubbard-Stratonovich transformation, the
expectation value of the current may be rewritten as
\begin{equation}\label{j-int-Seff}
\langle \, \mathbf{j} \, \rangle = \frac{ \int D \Delta \, D \bar
\Delta \, \langle \, \mathbf{j} \, \rangle_{\Delta \bar \Delta \phi} \,
e^{ - S_{\text{eff}} [\Delta, \bar \Delta, \phi]} }{ \int D \Delta \, D
\bar \Delta \, e^{-S_{\text{eff}} [\Delta, \bar \Delta, \phi ]} }.
\end{equation}
Here, $S_{\text{eff}}$ is the effective action for the pairing field
which is obtained by integrating out the fields
$\psi$,~\cite{SadeMelo-Randeria-Engelbrecht} and
\begin{equation}
\langle \, \mathbf{j} \, \rangle_{\Delta \bar \Delta \phi} = \frac{\int
D \psi \, D \bar \psi \, \mathbf{j} \, e^{- S_0 [\psi, \bar \psi, \phi]
+ \int d \tau \, d \mathbf{x} \, (\Delta \bar \psi_\uparrow \bar
\psi_\downarrow + \bar \Delta \psi_\downarrow \psi_\uparrow)}}{\int D
\psi \, D \bar \psi \, e^{- S_0 [\psi, \bar \psi, \phi] + \int d \tau
\, d \mathbf{x} \, (\Delta \bar \psi_\uparrow \bar \psi_\downarrow +
\bar \Delta \psi_\downarrow \psi_\uparrow)}} ,
\end{equation}
where $S_0$ is the part of the action $S$ which is quadratic in $\psi$.
The calculation of the contribution of superconducting fluctuations to
the current proceeds by applying a Gaussian approximation to
Eq.~(\ref{j-int-Seff}). More specifically, this involves expanding both
$S_{\text{eff}} [\Delta, \bar \Delta, \phi]$ and $\langle \mathbf{j}
\rangle_{\Delta \bar \Delta \phi}$ to second order in $\Delta$ and
$\bar \Delta$.

Consider first the calculation of the conductivity, in which case
$\phi$ is the electric field. In Eq.~(\ref{j-int-Seff}) the field
appears in two places, namely in $\langle \, \mathbf{j} \,
\rangle_{\Delta \bar \Delta \phi}$ and in the effective action
$S_{\text{eff}} [\Delta, \bar \Delta, \phi]$. The Aslamazov-Larkin
approximation involves keeping the field dependence in the effective
action only. The resulting expression is equivalent to a calculation
using a stochastic TDGL also done at Gaussian order: The quantity
$\langle \, \mathbf{j} \, \rangle_{\Delta \bar \Delta}$ is the current
associated with the order parameter configuration. The response of the
order parameter to the field is described by the effective action,
which is identical to the TDGL description. In particular note that in
the TDGL the electric field is coupled to linear order to the current
associated with the order parameter, as is described by the
Aslamazov-Larkin diagram. Not included in the Aslamazov-Larkin
approximation are the terms obtained by keeping the field in $\langle
\, \mathbf{j} \, \rangle_{\Delta \bar \Delta \phi}$. These describe
corrections to the normal state response modified by the presence of
the order parameter, and are the Maki-Thompson and density of states
corrections.

The situation for the transverse thermal response is somewhat different
than it is for the conductivity as we are considering the linear
response to two fields (since are considering the heat current response
to both electric and magnetic fields). The resulting diagrams are
presented in Fig.~\ref{fig-diagrams} (in most cases, the diagrams have
``mirror images'' which are not presented in the figure, but are of
course also taken into account). Before proceeding a word on
nomenclature: while the situation here is a bit different than in the
case of the conductivity, we will refer to the diagrams corresponding
to the TDGL contribution as the Aslamazov-Larkin diagrams; all other
diagrams will be collectively referred to as Maki-Thompson and density
of states diagrams, although they do not correspond to corrections to
the normal state transverse thermoelectric response only, as discussed
below.

The diagram in Fig.~\ref{fig-diagrams}(a) and its ``mirror image'' are
the Aslamazov-Larkin diagrams, which correspond to the contribution of
the stochastic TDGL. To obtain these diagrams, the electric and
magnetic fields are retained in the effective action in
Eq.~(\ref{j-int-Seff}) only, and the average over the current operator
(which in this case is the average over the heat current operator,
$\langle \, \mathbf{j}^\sQ \, \rangle_{\Delta \bar \Delta}$) gives the
heat current associated with the order parameter configuration.
Moreover, the motion of the order parameter described by these diagrams
is that of the TDGL, with the electric and magnetic fields coupled to
linear order to the electric currents associated with the order
parameter configuration. This correspondence will be revisited below:
In Sec.~\ref{sec-j} we discuss the heat current associated with the
motion of the order parameter $\langle \, \mathbf{j}^\sQ \,
\rangle_{\Delta \bar \Delta}$ and its connection to the heat current in
the TDGL, and in Sec.~\ref{sec-al} we calculate the Aslamazov-Larkin
diagrams for $\alpha_{xy}$ and find that they give the same
contribution as that found using a stochastic TDGL in
Ref.~\onlinecite{Ussishkin-Sondhi-Huse}.

The rest of the diagrams describe a variety of processes which involve
corrections to normal state properties, and may be understood along
similar lines. Diagrams (b) and (c) of Fig.~\ref{fig-diagrams} describe
a correction to the normal state thermoelectric response due to the
order parameter fluctuations, with Maki-Thompson [diagram (b)] and
density of state [diagram (c)] contributions, but with the order
parameter responding to linear order to the magnetic field. Likewise,
diagrams (d) and (e) describe the normal state response to a magnetic
field in the presence of superconducting fluctuations affected by the
electric field. Diagrams (f) and (g) describe the heat current
associated with an order parameter configuration (as in the
Aslamazov-Larkin diagrams), but with the dynamics of the order
parameter modified by a term not captured by the TDGL. (It is
interesting to note that in the microscopic picture there are
corrections to the TDGL in the order parameter motion in this case.)
Finally, diagrams (h)--(m) describe corrections to the normal state
transverse thermoelectric response $\alpha_{xy}$.

\section{Relation between current vertices}
\label{sec-j}

\begin{figure}
\begin{center}
  \includegraphics[width=3in]{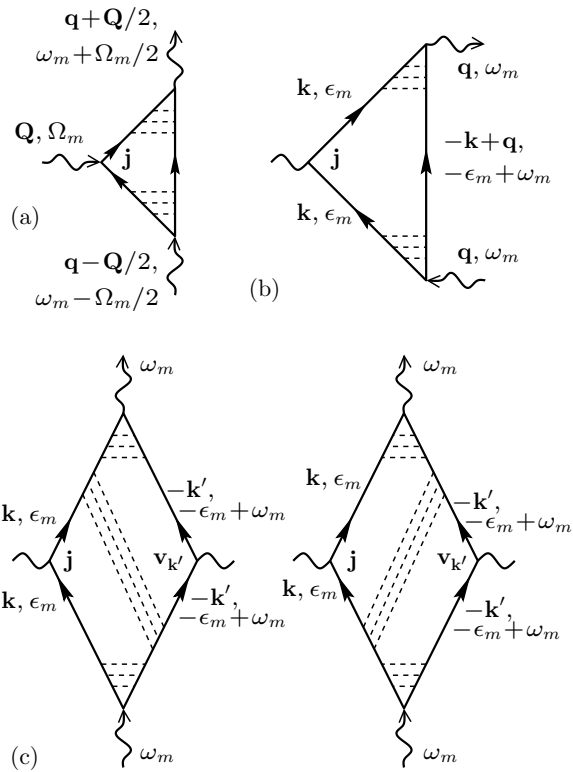}
\end{center}
\caption{\label{fig-vertex} Diagrams used in the text for evaluation of
the current vertices $\mathbf{J}^e$ and $\mathbf{J}^\sQ$ appearing in
the Aslamazov-Larkin diagrams. Lines with arrows are electronic Green
functions, and dashed lines denote disorder, shown here in the ladder
approximation. The microscopic current vertex $\mathbf{j}$ represents
either the electric current vertex $\mathbf{j}^e = - e
\mathbf{v}_\mathbf{k}$ or the heat current vertex $\mathbf{j}^\sQ = i
\epsilon_m \mathbf{v}_\mathbf{k}$.}
\end{figure}

In this section we consider the calculation of the triangular block
appearing in the Aslamazov-Larkin diagram [see
Fig.~\ref{fig-vertex}(a)]. In this diagram, the microscopic current
vertex $\mathbf{j}$ can be either an electric current vertex
$\mathbf{j}^e$ or a heat current vertex $\mathbf{j}^\sQ$. This diagram
corresponds to the current in the presence of an order parameter
configuration, and is thus directly related to the current which
appears in the TDGL. Accordingly, while the microscopic current vertex
is denoted with $\mathbf{j}$, we denote the current vertex presented by
the diagram in Fig.~\ref{fig-vertex}(a) with $\mathbf{J}$.

The result for the electric current vertex $\mathbf{J}^e$ is well known
(and is needed, e.g., for the conductivity~\cite{Aslamazov-Larkin}).
Our main concern here is with the heat current vertex $\mathbf{J}^\sQ$,
for which we establish the following result: At $Q = \Omega_m = 0$ [for
conventions regarding incoming and outgoing energies and momenta, see
Fig.~\ref{fig-vertex}(a)], the electric and heat current vertices are
related by
\begin{equation}\label{JQ-Je}
\mathbf{J}^\sQ = - \frac{i \omega_m}{2 e} \mathbf{J}^e .
\end{equation}
Heuristically this form is expected for a preformed pair of charge
$-2e$; but this is the BCS limit, for which an explicit calculation is
needed. Together with the known result for $\mathbf{J}^e$ [see
Eq.~(\ref{Je}) below] this allows the calculation of the
Aslamazov-Larkin diagrams in Sec.~\ref{sec-al}, as well as obtaining
the expression for the heat current in the
TDGL.~\cite{Ussishkin-Sondhi-Huse}

For completeness, we consider first the calculation of the electric
current vertex $\mathbf{J}^e$. The vertex $\mathbf{J}^e$ is needed in
the calculation of the Aslamazov-Larkin diagram to leading order in
external frequencies and momenta, and it is thus sufficient to set the
external frequencies to zero, $\Omega_m = \omega_m = 0$, and consider
linear order in the wavevectors $\mathbf{q}$ and $\mathbf{Q}$. There is
no linear term in $\mathbf{Q}$: indeed, it is straightforward to show
that $\mathbf{J}^e (q = \Omega_m = \omega_m = 0)$ is symmetric in
$\mathbf{Q}$.~\cite{note-Q=0} In the calculation of the electric
current vertex we thus set $Q = \Omega_m = \omega_m = 0$, and calculate
the vertex to linear order in $\mathbf{q}$, $\mathbf{J}^e
(\mathbf{q})$.

Setting $Q = \Omega_m = \omega_m = 0$ in $\mathbf{J}^e$, the diagram
for the electric current vertex is as in Fig.~\ref{fig-vertex}(b), but
with $\omega_m = 0$ (and $\mathbf{j} = \mathbf{j}^e = - e
\mathbf{v}_\mathbf{k}$).~\cite{Langer} The same diagram may be obtained
by inserting an electric current vertex in the diagram for $\Pi
(\mathbf{q}, \omega_m = 0)$, given in Fig.~\ref{fig-pi} (and correctly
accounting for the spin indices). Using the relation
\begin{equation}\label{vertex-in-G}
\nabla_\mathbf{k} G (\mathbf{k}, \epsilon_m) = \mathbf{v}_\mathbf{k} G
(\mathbf{k}, \epsilon_m)^2,
\end{equation}
where $G (\mathbf{k}, \epsilon_m)$ is the electron Green function, the
following result is obtained,
\begin{equation}\label{Je-Pi}
\mathbf{J}^e = - 2 e \nabla_\mathbf{q} \Pi (\mathbf{q}, \omega_m = 0).
\end{equation}
Note the appearance of the Cooper pair charge, $- 2 e$ (denoted below
as $e^*$). The expansion of Eq.~(\ref{Je-Pi}) to linear order in
$\mathbf{q}$ gives the result for the electric current
vertex,~\cite{Aslamazov-Larkin}
\begin{equation}\label{Je}
\mathbf{J}^e (\mathbf{q}) = - 4 e \eta \nu \mathbf{q}
\end{equation}
($\eta$ and $\nu$ were defined in Sec.~\ref{sec-fl}). The familiar
expression for the electric current in the TDGL (in the absence of
fields), follows from this result. More precisely, Eq.~(\ref{Je}) is
the electric current associated with a pairing field configuration
$\Delta = e^{i \mathbf{q} \cdot \mathbf{r}}$, using
\begin{equation}\label{TDGL-je}
\mathbf{J}^e = e^* \eta \nu \left[ \Delta^* (-i \nabla \Delta) +
\text{c.c.} \right] .
\end{equation}
[The conventional TDGL form, with $1 / 2 m^*$ replacing $\eta \nu$ (as
in, e.g., Ref.~\onlinecite{Skocpol-Tinkham}) is obtained after
rescaling the pairing field.]

We now reconsider the calculation of the electric current vertex, this
time with arbitrary $\omega_m$, as a first step towards establishing
Eq.~(\ref{JQ-Je}). Setting $Q = \Omega_m = 0$, the electric current
vertex $\mathbf{J}^e$ is presented in Fig.~\ref{fig-vertex}(b) (with
$\mathbf{j} = \mathbf{j}^e = - e \mathbf{v}_\mathbf{k}$). Using
Eq.~(\ref{vertex-in-G}), the expansion of $\mathbf{J}^e$ to linear
order in $\mathbf{q}$ is equivalent to introducing a second velocity
vertex. The vertex $\mathbf{J}^e$ is then written as $\mathbf{q}$ times
the contribution of square diagrams as in Fig.~\ref{fig-vertex}(c)
(with $\mathbf{j} = \mathbf{j}^e$). The following observation follows
from the structure of these diagrams: To linear order in $\mathbf{q}$,
but at arbitrary $\omega_m$, the electric current vertex has the
structure
\begin{equation}\label{Je-expansion}
\mathbf{J}^e = - e \, \mathbf{q} \sum_{\epsilon_m} f (\epsilon_m,
-\epsilon_m + \omega_m) \, ,
\end{equation}
where the function $f$ results from integration over all internal
momenta in the diagrams. The function $f$ depends only on the energy
variables appearing in the electronic Green functions, and running
along the two sides of the diagrams (we use here the fact that disorder
scattering is elastic). In addition, because of the symmetric structure
of the diagrams in Fig.~\ref{fig-vertex}(c), we have
\begin{equation}\label{f-symmetry}
f (\epsilon_m, -\epsilon_m + \omega_m) = f (- \epsilon_m + \omega_m,
\epsilon_m).
\end{equation}
The structure of the result for $\mathbf{J}^e$ as presented
Eqs.~(\ref{Je-expansion})--(\ref{f-symmetry}) is sufficient for
establishing Eq.~(\ref{JQ-Je}) to linear order in $\mathbf{q}$; it is
unnecessary to evaluate $f$ explicitly.

As with the electric current vertex, for the calculation of the
Aslamazov-Larkin diagram we only need the heat current vertex
$\mathbf{J}^\sQ$ to leading order in wavevectors and frequencies. As it
turns out [cf.\ Eq.~(\ref{JQ-Je})], this is one order higher than the
leading order in $\mathbf{J}^e$. By symmetry of the structure of the
diagrams for the current vertices, they are invariant under
$\mathbf{Q}, \Omega_m \rightarrow - \mathbf{Q}, -\Omega_m$. For the
expansion of $\mathbf{J}^\sQ$, this shows that there is no term linear
in $\mathbf{Q}$ or $\Omega_m$ alone, but does not exclude a term
proportional to $\mathbf{Q} \Omega_m$. In the calculation below, we use
the fact that in the Aslamazov-Larkin diagrams
[Fig.~\ref{fig-diagrams}(a) and its ``mirror image''] we need the heat
current vertex in the $\mathbf{\hat y}$ direction, perpendicular to
$\mathbf{Q} \parallel \mathbf{\hat x}$, and thus do not consider such a
term.

We thus proceed by setting $Q = \Omega_m = 0$. The heat current vertex
$\mathbf{J}^\sQ$ is then given in Fig.~\ref{fig-vertex}(b) (with
$\mathbf{j} = \mathbf{j}^\sQ = i \epsilon_m
\mathbf{v}_\mathbf{k}$).~\cite{Langer} As in the case of the electric
current vertex $\mathbf{J}^e$, the expansion of $\mathbf{J}^\sQ$ to
linear order in $\mathbf{q}$ amounts to the introduction of a second
velocity vertex, resulting in square diagrams as in
Fig.~\ref{fig-vertex}(c) (with $\mathbf{j} = \mathbf{j}^\sQ$). It
follows from the structure of these diagrams that to linear order in
$\mathbf{q}$ (and at arbitrary $\omega_m$) the heat current vertex has
the structure
\begin{equation}\label{JQ-expansion}
\mathbf{J}^\sQ = \mathbf{q} \sum_{\epsilon_m} i \epsilon_m f
(\epsilon_m, -\epsilon_m + \omega_m) ,
\end{equation}
where the function $f$ results from integration over all internal
momenta in the diagrams. The important point here is that the function
$f$ that appears in Eq.~(\ref{Je-expansion}) is \emph{identical} to the
one that appears in Eq.~(\ref{JQ-expansion}). Eq.~(\ref{JQ-expansion})
may be rewritten as
\begin{eqnarray}
\mathbf{J}^\sQ & = & \mathbf{q} \sum_{\epsilon_m} \left( i
\epsilon_m - \frac{i \omega_m}{2} \right) f (\epsilon_m,
-\epsilon_m + \omega_m) \nonumber \\
& + &  \mathbf{q} \sum_{\epsilon_m} \frac{i \omega_m}{2} f (\epsilon_m,
-\epsilon_m + \omega_m).
\end{eqnarray}
Here, the first term on the right hand side can be shown to vanish
using Eq.~(\ref{f-symmetry}). On comparing the second term with
Eq.~(\ref{Je-expansion}), we find the relation between the current
vertices, Eq.~(\ref{JQ-Je}).

We note that to this order of the calculation $\mathbf{J}^\sQ$ does not
have a branch cut after analytic continuation of $i \omega_m$ to the
$\omega$-plane. After analytic continuation, we thus have to linear
order in $\mathbf{q}$ and $\omega$
\begin{equation}\label{JQ}
\mathbf{J}^\sQ (\mathbf{q}, \omega) = - \frac{\omega}{2 e}
\mathbf{J}^e (\mathbf{q}) = 2 \eta \nu \omega \mathbf{q}.
\end{equation}
This result is used below in the calculation of the Aslamazov-Larkin
diagrams, and may be used to obtain the heat current in the
TDGL~\cite{Ussishkin-Sondhi-Huse} (again, note the appearance of the
Cooper pair charge, $-2e$). To be precise, Eq.~(\ref{JQ}) is the heat
current associated with a pairing field configuration $\Delta = e^{i
\mathbf{q} \cdot \mathbf{r} - i \omega t}$, using [cf.
Eq.~(\ref{TDGL-je})]
\begin{equation}
\mathbf{J}^\sQ = - \eta \nu \left[ \frac{\partial \Delta^*}{\partial t}
\nabla \Delta + \text{c.c.} \right].
\end{equation}

Previously, the heat current vertex $\mathbf{J}^\sQ$ was considered by
several authors, beginning with the work of Caroli and
Maki.~\cite{Caroli-Maki} However, Eqs.~(\ref{JQ-Je}) and (\ref{JQ}) do
not appear to have been obtained previously with the correct
factor.~\cite{note-CM} The same result may of course be obtained by an
explicit (but more cumbersome) calculation of the heat current vertex
$\mathbf{J}^\sQ$.~\cite{Reizer-Sergeev} However, in addition to their
being more straightforward, the arguments presented in this section
have the advantage of being very general, applicable to arbitrary
disorder strength and range. [Note that the important ingredient used
to obtain Eqs.~(\ref{Je-expansion}), (\ref{f-symmetry}), and
(\ref{JQ-expansion}) is just the absence of inelastic scattering.]
Finally, we have shown here that Eq.~(\ref{JQ-Je}) holds to linear
order in $\mathbf{q}$; we note that the argument may be extended to
higher orders in $\mathbf{q}$ as well.

\section{Calculation of the Aslamazov-Larkin diagrams}
\label{sec-al}

\begin{figure}
\begin{center}
  \includegraphics[width=3in]{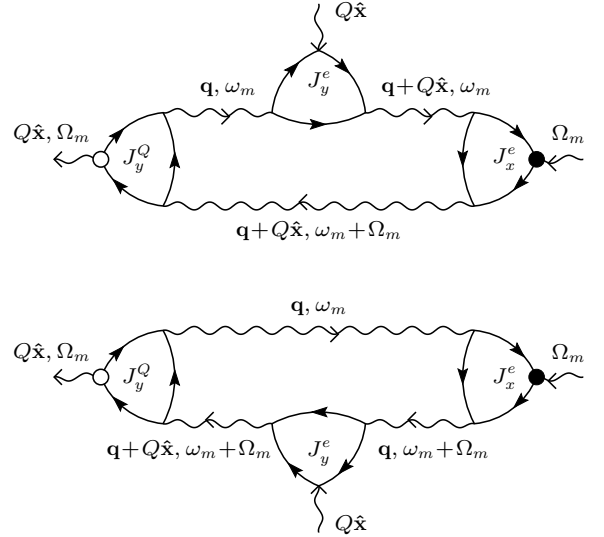}
\end{center}
\caption{\label{fig-al} Aslamazov-Larkin diagrams contributing to
$j_y^\sQ / EB$ [See Eqs.~(\ref{Kubo-jyQ-EB})--(\ref{jjj-correlator})].
The wavy lines correspond to the fluctuation propagator $L$; electric
and heat current vertices are indicated in the figure.}
\end{figure}

\begin{figure}
\begin{center}
  \includegraphics[width=3in]{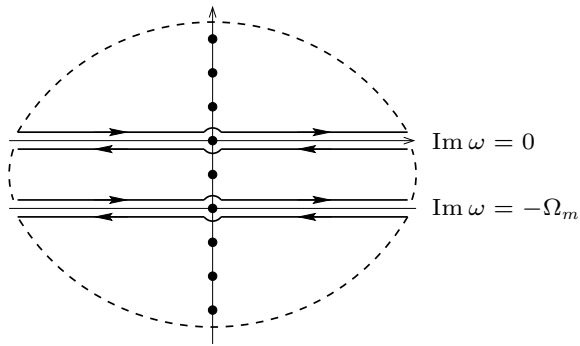}
\end{center}
\caption{\label{fig-contour} Contour in the complex $\omega$ plane used
for expressing the sum over Matsubara frequencies $\omega_m$ in
Eq.~(\ref{JJJ-sum}) as an integral, leading to Eq.~(\ref{JJJ-int}). The
contour runs along the cuts of $L (\mathbf{q}, \omega)$ and $L
(\mathbf{q}, \omega + i \Omega_m)$, and is closed at a distance from
the origin which is taken to infinity. The dots are the poles of
$n(\omega)$ at $i \omega_m$.}
\end{figure}

In this section we calculate the Aslamazov-Larkin contribution to
$\alpha_{xy}$. The starting point is the expression for the
Aslamazov-Larkin diagrams depicted in Fig.~\ref{fig-al}. To leading
order in momentum and energy, the current vertices depend only on
momentum and energy flowing from one fluctuation propagator to the
other, as in Eqs.~(\ref{Je}) and (\ref{JQ}). It follows that the
Aslamazov-Larkin contribution to the current correlator $\Lambda$ [see
Eq.~(\ref{jjj-correlator})] is given by
\begin{widetext}
\begin{eqnarray}\label{JJJ-sum}
\Lambda^\sAL (Q, i \Omega_m) & = & - \frac{1}{\beta} \sum_{\omega_m}
\int \frac{d \mathbf{q}}{(2 \pi)^d}
\\
\nonumber & \times & \Big\{ J_x^e (q_x + Q) \, J_y^e (q_y) \, J_y^\sQ (q_y,
i \omega_m + i \Omega_m / 2) \, L (\mathbf{q}, i \omega_m) \, L (\mathbf{q}
+ Q \mathbf{\hat x}, i \omega_m) \, L (\mathbf{q} + Q \mathbf{\hat x}, i
\omega_m + i \Omega_m)
\\
\nonumber & & + \; J_x^e (q_x) \, J_y^e (q_y) \, J_y^\sQ (q_y, i \omega_m +
i \Omega_m / 2) \, L (\mathbf{q}, i \omega_m) \, L (\mathbf{q}, i \omega_m
+ i \Omega_m) \, L (\mathbf{q} + Q \mathbf{\hat x}, i \omega_m + i
\Omega_m) \Big\} .
\end{eqnarray}
Following the standard procedure, the sum over Matsubara frequencies
may be expressed as an integral over the contour presented in
Fig.~\ref{fig-contour}. The resulting expression, after the analytic
continuation $i \Omega_m \rightarrow \Omega + i 0$, is given by
\begin{eqnarray}\label{JJJ-int}
\Lambda^\sAL (Q, \Omega) & = & - \frac{1}{\pi} \int_{-\infty}^\infty d
\omega \, n (\omega) \int \frac{d \mathbf{q}}{(2 \pi)^d}
\\
\nonumber & \times & \Bigg( J_x^e (q_x + Q) \, J_y^e (q_y) \, \bigg\{
J_y^\sQ (q_y, \omega + \Omega / 2) \, L^\sR (\mathbf{q} + Q \mathbf{\hat
x}, \omega + \Omega) \, \text{Im} \Big[ L^\sR (\mathbf{q}, \omega) L^\sR
(\mathbf{q} + Q \mathbf{\hat x}, \omega) \Big]
\\
\nonumber & & \qquad \qquad \qquad \qquad \;\;\; + \; J_y^\sQ (q_y, \omega
- \Omega / 2) \, L^\sA (\mathbf{q}, \omega - \Omega) \, L^\sA (\mathbf{q} +
Q \mathbf{\hat x}, \omega - \Omega) \, \text{Im} \Big[ L^\sR (\mathbf{q} +
Q \mathbf{\hat x}, \omega) \Big] \bigg\}
\\
\nonumber & & \;\; + \;  J_x^e (q_x) \, J_y^e (q_y) \, \bigg\{ J_y^\sQ
(q_y, \omega + \Omega / 2) \, L^\sR (\mathbf{q}, \omega + \Omega) \, L^\sR
(\mathbf{q} + Q \mathbf{\hat x}, \omega + \Omega) \, \text{Im} \Big[ L^\sR
(\mathbf{q}, \omega) \Big]
\\
\nonumber & & \qquad \qquad \qquad \;\;\;\;\;\; + \; J_y^\sQ (q_y, \omega -
\Omega / 2) \, L^\sA (\mathbf{q}, \omega - \Omega) \, \text{Im} \Big[ L^\sR
(\mathbf{q}, \omega) L^\sR (\mathbf{q} + Q \mathbf{\hat x}, \omega) \Big]
\bigg\} \Bigg) .
\end{eqnarray}
\end{widetext}
Here, $n(\omega) = \frac{1}{2} \coth (\omega / 2 T)$, and $L^\sR
(\mathbf{q}, \omega)$ and $L^\sA (\mathbf{q}, \omega)$ are the analytic
continuation of $L (\mathbf{q}, i \omega_m)$ on the two sides of the
cut at $\text{Im} \, \omega = 0$. The main contribution to the
integrals is from small wavevectors and frequencies; to leading order
in $T - T_c$, $n(\omega) \approx T / \omega$, and the fluctuation
propagator $L$ and current vertices $\mathbf{J}^e$ and $\mathbf{J}^\sQ$
are given by Eqs.~(\ref{L}), (\ref{Je}), and (\ref{JQ}) respectively.
Next, the expression is expanded to linear order in $Q$ and $\Omega$,
the integrals are calculated, and using Eq.~(\ref{Kubo-jyQ-EB}), we
obtain for two and three dimensions,
\begin{equation}\label{jQ-EB}
\frac{j_y^\sQ}{E} = \left\{
\begin{array}{lll}
\displaystyle - \frac{e^2 T B}{2 \pi c} \, \frac{\eta}{\epsilon} &
\qquad & \text{for 2D,}
\\
\\
\displaystyle - \frac{e^2 T B}{4 \pi c} \, \sqrt{\frac{\eta}{\epsilon}}
& & \text{for 3D.}
\end{array}
\right.
\end{equation}
With the identification of the superconducting coherence length $\xi(T)
= \sqrt{\eta / \epsilon}$, this result is identical to the one obtained
by considering the Gaussian fluctuations in a stochastic
TDGL.~\cite{Ussishkin-Sondhi-Huse,Ullah-Dorsey}

As discussed in Sec.~\ref{sec-Kubo}, it is necessary to subtract the
magnetization current $\mathbf{j}_{\text{mag}}^\sQ = c \mathbf{M}
\times \mathbf{E}$ from this result to obtain the correct transport
response. The corresponding contribution of superconducting
fluctuations to the magnetization is given
by~\cite{Skocpol-Tinkham,Larkin-Varlamov}
\begin{equation}\label{M}
\mathbf{M} = \left\{
\begin{array}{lll}
\displaystyle - \frac{e^2 T B}{3 \pi c^2} \, \frac{\eta}{\epsilon} &
\qquad & \text{for 2D,}
\\
\\
\displaystyle - \frac{e^2 T B}{6 \pi c^2} \,
\sqrt{\frac{\eta}{\epsilon}} & & \text{for 3D.}
\end{array}
\right.
\end{equation}
We note that in the Aslamazov-Larkin calculation the magnetization
currents contribute two thirds of the total current in both two and
three dimensions.

The final result for the Aslamazov-Larkin contribution to $\alpha_{xy}$
is obtained after the subtraction of the magnetization currents. The
result is given in Eq.~(\ref{alpha-xy-result}), where we introduce back
$\hbar$, and present the result in terms of the coherence length,
$\xi(T) = \sqrt{\eta / \epsilon}$.

\section{Maki-Thompson and density of states terms}
\label{sec-mt}

In this section we consider the Maki-Thompson and density of states
diagrams [diagrams (b)--(m) in Fig.~\ref{fig-diagrams}]. We show here
that these terms are less divergent than the Aslamazov-Larkin diagrams
as $T \rightarrow T_c$. In addition a similar conclusion can be drawn
for the magnetization. We thus find that in the microscopic calculation
of $\alpha_{xy}$ the Aslamazov-Larkin result,
Eq.~(\ref{alpha-xy-result}), is the most divergent contribution as $T
\rightarrow T_c$.

We note that a similar result may or may not hold for different
transport coefficients, and each case must be examined separately.
Indeed, in the case of the conductivity in three dimensions the
Maki-Thompson contribution has the same divergence as the
Aslamazov-Larkin contribution.~\cite{Maki} In two dimensions the
Maki-Thompson diagram diverges at any temperature in a naive
calculation, a divergence which is regularized by introducing a pair
breaking mechanism.~\cite{Thompson,Patton,Keller-Korenman} After this
regularization, the Maki-Thompson contribution has a form which is
somewhat different than the power law of the Aslamazov-Larkin term, and
its ultimate divergence is only logarithmic. Nevertheless, this is an
important microscopic contribution to the conductivity (except very
close to $T_c$). In contrast, a different situation may hold for other
transport properties. For example, Niven and Smith have recently shown
that the contribution of the Maki-Thompson and density of states terms
for the thermal conductivity does not diverge at
$T_c$.~\cite{Niven-Smith}

The method that we use here to obtain our result is that of power
counting, applied to each of the diagrams independently. We thus avoid
the explicit calculation of the diagrams, which would be needed if
sub-leading terms are desired. In each of these diagrams, after each of
the electronic blocks is calculated, the structure that remains is that
of an integral over momentum and energy, with the integrand being
composed of fluctuation propagators and electronic blocks. We apply
power counting arguments to this integral to find the dependence of
each diagram on the reduced temperature $\epsilon$. (This procedure
does not exclude the possibility that the coefficient of this power is
actually zero and that the power of the diagram is therefore lower, nor
does it exclude the possibility that the diagram is identically zero.)

For the purpose of clarification, we begin by considering the
Aslamazov-Larkin diagram [Fig.~\ref{fig-diagrams}(a)], which was
calculated explicitly in Sec.~\ref{sec-al}. The power counting is thus
applied to Eq.~(\ref{JJJ-int}); we now count powers of $\epsilon$ in
the integral explicitly. In the integrand, there are three fluctuation
propagators $L$ (contributing a power $\epsilon^{-1}$ each), two
electric current vertices $\mathbf{J}^e$ ($\epsilon^{1/2}$ each), and
one heat current vertex $\mathbf{J}^\sQ$ ($\epsilon^{3/2}$). To obtain
$j_y^\sQ / EB$ [see Eq.~(\ref{Kubo-jyQ-EB})], the integral is expanded
in external frequency $\Omega$ ($\epsilon^{-1}$) and external
wavevector $Q$ ($\epsilon^{-1/2}$). The integration over momentum gives
another $\epsilon^{d/2}$, while there is no contribution associated
with the integration over energy [because of the $n(\omega)$ factor].
Accounting for all contributions, we obtain a divergence of
$\epsilon^{d/2 - 2}$. Similar arguments give an identical result for
the divergence of the magnetization, giving $\alpha_{xy}^\sAL \propto
\epsilon^{d/2 - 2}$, in agreement with our exact calculation,
Eq.~(\ref{alpha-xy-result}).

We consider next diagrams (b)--(g) in Fig.~\ref{fig-diagrams}. In these
diagrams, the number of fluctuation propagators is one less than in the
Aslamazov-Larkin diagram. If any of these diagrams is to be as
divergent as the Aslamazov-Larkin diagram, then this loss of a power of
$\epsilon^{-1}$ must be compensated (as it is in the case of the
Maki-Thompson diagram in the conductivity). A power of $\epsilon^{-1}$
is regained when considering the electronic block of the diagram which
has two microscopic current vertices in it, provided they are in the
same direction. Here, this will only occur for diagrams~(d) and~(e), in
which the two vertices which are in the same block are $j_y^e$ and
$j_y^\sQ$. Indeed, in the Aslamazov-Larkin diagram, the vertices
$J_y^e$ and $J_y^\sQ$ contribute $\epsilon^2$. In diagrams~(d) and~(e),
the block containing the vertices $j_y^e$ and $j_y^\sQ$ contributes
only one power of $\epsilon$.~\cite{note-MT} Moreover, for these two
diagrams to diverge as the Aslamazov-Larkin diagram, the expansion in
external wavevector $Q$ and external frequency $\Omega$ should give
powers of $\epsilon^{-1/2}$ and $\epsilon^{-1}$ respectively, as it
does for the Aslamazov-Larkin diagram. The expansion in $\Omega$ indeed
gives a power of $\epsilon^{-1}$, since the external frequency appears
explicitly in the fluctuation propagator (and also due to a diffusive
pole as in the case of the Maki-Thompson conductivity diagram).
However, the important point in this analysis is that in diagrams (d)
and (e) the expansion in external momentum $Q$ does not gain a power of
$\epsilon^{-1/2}$, but instead accounts for $\epsilon^{1/2}$ in the
power counting. The reason for this is that the external momentum,
which flows from the $j_y^e$ vertex to the $j_y^\sQ$ vertex, does not
flow through the fluctuation propagators of the diagram. The expansion
in external momentum is thus limited to the electron block which
includes these two vertices, where it is straightforward to check that
expansion in the external momentum leads to a power of
$\epsilon^{1/2}$. Finally, diagrams (h)--(m) involve only one
fluctuation propagator and will clearly be less divergent than the
Aslamazov-Larkin term.

We have demonstrated by power counting arguments that while the
Aslamazov-Larkin diagram, Fig.~\ref{fig-diagrams}(a), diverges as
$\epsilon^{d/2 - 2}$, all other diagrams in Fig.~\ref{fig-diagrams} are
less divergent as $T \rightarrow T_c$. Similar arguments hold for the
fluctuation contribution to the magnetization, and hence for
$\alpha_{xy}$. In view of this conclusion we do not calculate the
Maki-Thompson and density of diagrams explicitly. [We note that
diagrams (b)--(g) in Fig.~\ref{fig-diagrams}, as well as the
Maki-Thompson and density of states diagrams in the calculation of the
magnetization, will have a logarithmic divergence in two dimensions as
$T \rightarrow T_c$.]

\section{Conclusions}
\label{sec-summary}

The main result of this paper is that the leading contribution in the
microscopic calculation of $\alpha_{xy}$ arises from the
Aslamazov-Larkin diagrams, which correspond to the contribution of
Gaussian fluctuations in the stochastic TDGL. The Maki-Thompson and
density of state terms are less divergent as $T \rightarrow T_c$. In
concluding this paper, we comment on several aspects of this result.

It is well known that in calculating the contribution of
superconducting fluctuations to the conductivity, the Aslamazov-Larkin
contribution corresponds to the Gaussian contribution of a stochastic
TDGL.~\cite{Skocpol-Tinkham,Larkin-Varlamov} In establishing the
correct form for the heat current vertex in Sec.~\ref{sec-j}, we verify
this correspondence also for thermal transport.

There are two directions in which to extend the calculation beyond this
approximation: by considering the additional microscopic contributions
(as we did in this paper), or by going beyond the Gaussian
approximation in the stochastic TDGL (cf.\
Ref.~\onlinecite{Ussishkin-Sondhi-Huse}). We would like to emphasize
that these approaches are of a very different nature, and their regime
of validity is also different.

The stochastic TDGL is traditionally understood as the model for the
critical dynamics of a superconductor (model A in the classification of
Hohenberg and Halperin~\cite{Hohenberg-Halperin}). As such, the TDGL
should give the relevant contribution as the temperature approaches
$T_c$ in the critical regime (which for low-temperature superconductors
is very narrow, as expressed by the Ginzburg
criterion~\cite{Larkin-Varlamov}). Additional microscopic terms become
irrelevant in this regime.

On the other hand, further away from $T_c$ in the region where the
microscopic calculation is valid, additional microscopic contributions
may arise (as they do for the conductivity). To reiterate, for the
transverse thermoelectric response we find that they are less divergent
then the Aslamazov-Larkin contribution as $T \rightarrow T_c$. Not
investigated here is the possibility that these normal state
corrections vanish in the case of particle-hole symmetry, as does the
Drude result for $\alpha_{xy}$ (to emphasize, a result not required by
symmetry).

To connect the microscopic approach with the critical dynamics, one may
expect that in the microscopic approximation as the temperature is
lowered, the Maki-Thompson and density of states terms would become
less important as the behavior becomes governed by the stochastic TDGL.
For $\alpha_{xy}$ this clearly occurs in the microscopic
calculation.~\cite{note-Patton} In conclusion, this work provides
further justification for using the TDGL also as the temperature
increases away from $T_c$ into the Gaussian regime (which is the
approach taken in Ref.~\onlinecite{Ussishkin-Sondhi-Huse}).

Interest in the Nernst signal has grown recently due to the
measurements in high-temperature superconductors, where the
fluctuations signal can be observed well above
$T_c$.~\cite{Ong-expt,Capan-etal} On the other hand, the contribution
of superconducting fluctuations to the Nernst signal is yet to be
observed in a low-temperature superconductor, for which the BCS
microscopics considered in this paper are applicable. As discussed in
the Introduction, we expect the fluctuation tail to be observable in
the Nernst signal of a suitably chosen superconducting thin film. It
would certainly be of interest to verify this prediction
experimentally.

\begin{acknowledgments}

I thank Shivaji Sondhi and David Huse, my collaborators on ongoing
related work on thermal transport and the Nernst effect in
superconductors, for numerous illuminating discussions. I would also
like to acknowledge discussions with Vadim Oganesyan and Austen
Lamacraft.

\end{acknowledgments}


\begin{thebibliography}{29}
\bibitem{note-Q=0}
Actually, for our purposes one could also set $Q = 0$ from the way the
Aslamazov-Larkin diagram is constructed. In the conductivity
calculation~\cite{Aslamazov-Larkin} the two current vertices have $Q =
0$. Here, for $\alpha_{xy}$, the Aslamazov-Larkin diagram of
Fig.~\ref{fig-diagrams}(a) has two electric current vertices, one with
$Q = 0$ and one with a vertex in the $\mathbf{\hat y}$ direction, but
with $\mathbf{Q} \parallel \mathbf{\hat x}$.
\bibitem{note-MT}
Here, and in the rest of this section, we use the assumption of
particle-hole symmetry in obtaining the powers of electronic blocks
containing $j_y^\sQ$.
\bibitem{note-Patton} A similar observation follows also for the microscopic
calculation of the conductivity, albeit in a more subtle manner: The
Maki-Thompson term is regularized by pair breaking, and becomes weaker
than the Aslamazov-Larkin term as $T \rightarrow T_c$. This is true
even in the absence of explicit pair
breaking.~\cite{Patton,Keller-Korenman} In this case, however, not too
close to $T_c$ the Maki-Thompson term is an important contribution to
the microscopics.
\end{thebibliography}

\end{document}